\documentclass[aps, prd, twocolumn, superscriptaddress, showpacs, nofootinbib]{revtex4-1}
\usepackage[utf8]{inputenc}
\usepackage[T1]{fontenc}
\usepackage{ae,aecompl} 
\usepackage{graphicx}
\usepackage{amsmath}
\usepackage{color}
\usepackage{amssymb}
\usepackage{latexsym}
\usepackage{wasysym}
\usepackage{psfrag}
\usepackage{ifthen}
\usepackage[citecolor=blue,colorlinks=true]{hyperref}
\usepackage{longtable}
\usepackage{float}
\usepackage[utf8]{inputenc}
\usepackage{lineno}
\usepackage{units}
\usepackage{multirow}
\usepackage{orcidlink}
\def\cqg{Classical Quantum Gravity}   
\def\apjl{Astrophys.~J.~Lett}
\def\apjs{Astrophysical Journal, Supplement}

\def\jcap{Journal of Cosmology and Astroparticle Physics}
\def\mnras{Monthly Notices of the RAS}

\newcommand{\note}[1]{{\color{black}#1}}

\newcommand{\revision}[1]{{\color{black}#1}}
\newcommand{\revisions}[1]{{\color{black}#1}}

\begin{abstract}

\revision{Einstein Telescope (ET) is a third-generation gravitational-wave (GW) detector project which aims to reach a sensitivity $\sim10$ times larger than current GW detectors such as Advanced LIGO and Advanced Virgo.}
The high rate of GW signals expected in the data will pose several data analysis challenges, such as the ability to disentangle overlapping signals, or the computational resources required to treat all the candidate events.
We explore the behaviour and the performances of \revision{\texttt{PySTAMPAS}, }a data analysis pipeline designed to search for unmodeled GW signals with duration \unit[$\sim1-1000$]{s} on a mock dataset that consists of $1$ month of data following ET design sensitivity on top of which is added a realistic distribution of compact binary coalescence (CBC) signals.
Unmodeled searches are intrinsically less sensitive to CBC signals than template-based searches, but are computationally cheaper and more robust to uncertainties in the waveforms.
This search recovers $38\%$ of the total number of injected binary black hole (BBH) signals, including $89\%$ of the systems with a total mass above \unit[$100$]{M$_\odot$}, as well as the majority of binary neutron star (BNS) signals closer than \unit[$850$]{Mpc} ($z=0.17$). It is also able to estimate the chirp mass of the recovered BNS with an average precision of $1.3 \%$.
Therefore, we show that this unmodeled search is able to detect a substantial amount of CBC events at a relatively low computational cost, which makes it interesting for low-latency analyses and independent validation of detections made by matched filtering pipelines.
We also find that the presence of many CBC signals only marginally impacts the sensitivity of the search to other kinds of unmodeled long-duration transient signals, by $\sim 3\%$ in average.
\end{abstract}

\begin{document}
\title{A weakly-modeled search for compact binary coalescences in Einstein Telescope}
\author{Adrian Macquet\,\orcidlink{0000-0001-5955-6415}}
\affiliation{Université Paris-Saclay, CNRS/IN2P3, IJCLab, 91405 Orsay, France}
\author{Tito \surname{Dal Canton}\,\orcidlink{0000-0001-5078-9044}}
\affiliation{Université Paris-Saclay, CNRS/IN2P3, IJCLab, 91405 Orsay, France}
\author{Tania Regimbau}
\affiliation{Univ. Savoie Mont Blanc, CNRS, Laboratoire d’Annecy de Physique des Particules - IN2P3, F-74000 Annecy, France}
\date[\relax]{compiled \today}

\maketitle

\section{Introduction}


Third-generation gravitational-wave (GW) detectors such as Einstein Telescope (ET) \cite{ET2010} and Cosmic Explorer \cite{CE2019, CE2021} are expected to become operational in the mid-$2030s$.
They aim to improve the sensitivity by an order of magnitude compared to the current generation of advanced detectors (Advanced LIGO \cite{2015CQGra..32g4001L}, Advanced Virgo \cite{2015CQGra..32b4001A} and KAGRA \cite{2019NatAs...3...35K}, commonly referred to as LVK), and to extend the sensitive frequency band at both low and high frequencies.
In its current planed design, ET will be made of three nested detectors forming an equilateral triangle with \unit[$10$]{km} sides. Each detector will consist \revision{of} two dual-recycled Fabry-Perot-Michelson interferometers tuned to be sensitive to high and low frequencies respectively, and the whole infrastructure will be built underground to limit the impact of seismic noise.

At its design sensitivity, ET should be able to detect most of the binary black holes mergers (BBH) with total mass in the range \unit[$10^1-10^3$]{M$_\odot$} up to a redshift $z=20$, and binary neutron stars (BNS) up to $z=2$ \cite{2020JCAP...03..050M}. The expected rate of detectable compact binary coalescences (CBC) lies in the hundreds to thousands per day.
In addition to CBC, which are expected to remain the dominant source in ET, other sources of GW could be present in the frequency band, such as continuous emission from isolated neutron stars \cite{2019Univ....5..217S}, stochastic GW backgrounds \cite{2014PhRvD..89h4046R}, and other transient sources not from CBC origin, like core-collapse supernovae \cite{2021PhRvD.104j2002S} or magnetar flares \cite{2001MNRAS.327..639I, 2011PhRvD..83j4014C}. An extensive description of the scientific targets of ET can be found in \cite{2023JCAP...07..068B}.

The sheer \revision{number} of signals in the data, a fraction of which with signal-to-noise ratio (SNR) above $20$ \cite{FindChirp}, will introduce new data analysis challenges.
Because of the extended sensitivity at low frequency, CBC signals will be present for a longer time in the frequency band of the detectors, up to several hours or days in the case of BNS \cite{2009PhRvD..79f2002R}. A fraction of these signals will overlap each other, which may affect the ability of data analysis pipelines to detect and to estimate the parameters of such signals \cite{2012PhRvD..86l2001R, 2023MNRAS.523.1699J, 2024PhRvD.109h4015J, 2022PhRvD.105j4016P, 2021PhRvD.104h4039R, 2021PhRvD.104d4003S, 2021MNRAS.507.5069A, 2022PhRvD.106j4045R}.
The sum of all CBC events that are not individually resolvable will also create a confusion background that will complicate the observation of a stochastic GW background of cosmological origin, which is one of the big targets of GW astronomy \cite{2011RAA....11..369R, 2014PhRvD..89h4046R, 2018CQGra..35p3001C, 2019RPPh...82a6903C, 2023PrPNP.12804003V} (see \cite{2021PhRvD.104b2004A} for the most recent constraints provided by LVK). Methods to subtract this background are therefore being developed \cite{2020PhRvD.102b4051S, 2023PhRvD.108f4040Z, 2023PhRvD.107f4048Z}.
Finally, the computational resources used to analyze the data will need to be scaled up to account for the larger bandwidth of the detector and a rate of events orders of magnitude higher than with current detectors \cite{Lenon2021,2023arXiv231211103B}.
In the context of multi-messenger astronomy, it is also crucial to provide rapid signal detections and estimations of the spatial position of the source, in order to trigger space- and Earth-based electromagnetic observations of the same source as soon as possible \cite{GW170817MMA}. Therefore, robust and rapid search algorithms will be needed to cope with the rate of detectable signals, such as the one proposed in \cite{Miller:2023rnn}.

Searches for GW transients are generally separated into two main categories, modeled and unmodeled.
When the waveform is well modeled and depends on a relatively small number of parameters, as it is generally the case for CBC, the signal can be searched for nearly-optimally using matched filtering \cite{FindChirp}. 
However, when the waveform is characterized by a too large number of parameters, or the GW emission process is not entirely understood or too complex to provide a precise parametric waveform model, matched filtering techniques cannot be used and the search must rely on signal-agnostic detection algorithms.
Transient GW signals that are not searched using matched filtering techniques are commonly referred to as \textit{bursts}. 
They can be CBC signals that lie outside the parameter space scanned by template-based searches, such as systems with high eccentricity, strong precession or matter effects \cite{2018PhRvD..97b4031H}, or other sources such as core-collapse supernovae, oscillations in isolated neutron stars, and accretion disk instabilities around black holes \cite{ADI}.
Most unmodeled GW search algorithms consist in looking for an excess of energy in some time-frequency representation of the data, and using cross-correlation between two or more detectors to discriminate true GW events from noise.
By design, these searches are less sensitive than template-based ones for the detection of CBC events. However, they often require much less computational resources, which could prove important for ET when thousands of events are expected per day. They are also not affected by uncertainties in waveform models, and can also cover parts of the parameter space that are not included in standard CBC searches. 

\revision{\texttt{PySTAMPAS} \cite{PySTAMPAS} is a data analysis pipeline originally designed to search for long-duration (\unit[$\sim 1-1000$]{s}) GW bursts in LVK data.}
In this work, we investigate whether \revision{this specific} search algorithm \revision{could also} be used to detect CBC signals in ET and to provide an estimation of the chirp mass, and we study the impact of the CBC foreground on the detectability of other types of unmodeled transients. \revision{The choice of a search algorithm that targets signals with duration $> 1$ s is motivated by the extended duration of CBC signals due to the improved sensitivity of ET at low frequency.} 
\revision{This work takes place in the context of the latest ET mock data challenge (MDC) which aims to provide a common dataset to investigate how current data analysis techniques would perform in the context of ET, identify the science cases, and test new data analysis and parameter estimation methods.}
In this paper, we analyse this MDC that consists of one month of data simulating ET noise and a realistic distribution of CBC events, with \texttt{PySTAMPAS}.
The paper is organized as follows. In Sec. \ref{sec:method}, we present the dataset and describe the search algorithm we use to analyze it. Results of these analyzes are reported in Sec. \ref{sec:results}. In Sec. \ref{sec:unmodeled}, we study the ability of the pipeline to detect non-CBC transient signals in the presence of a CBC foreground. We discuss the results obtained and the takeways of this study in Sec. \ref{sec:conclusion}

\section{Dataset and methodology}
\label{sec:method}
\subsection{Description of the dataset}

This MDC contains $31$ days of data simulating the output of the three V-shaped nested interferometers forming ET in the triangular configuration. 
For each detector, the data are the sum of colored Gaussian noise and the detector's response to a distribution of GW signals from CBC sources.
The dataset was generated with the with the \texttt{MDC\_Generation} pipeline \cite{MDC_Generation}, a code \revision{already} used for past MDC \cite{2012PhRvD..86l2001R, 2014PhRvD..89h4046R, 2016PhRvD..93b4018M}. To compute the response of the three detectors to GW signals, it is assumed that they are located at the current EGO site near Cascina in Italy.
The PSD of the noise is the reference one defined in \cite{2023JCAP...07..068B} for \unit[$10$]{km} arms in the xylophone configuration (combining high and low-frequency interferometers). The noise is purely Gaussian, colored with the PSD, and uncorrelated between the detectors.

In total, $69781$ CBC signals are present in the data, split into $3$ source classes: $61031$ BNS, $6725$ BBH, and $2025$ neutron star-black hole mergers (NSBH).
The population parameters of BNS and NSBH follow \cite{2021MNRAS.502.4877S}, and the BBH \cite{2022MNRAS.511.5797M}. They are compatible with the merger rates estimated by LVK observations (GWTC-3 catalog) \cite{2023PhRvX..13d1039A}. Assuming a Poisson process, the time interval between successive coalescence times follow an exponential distribution $P(\tau) = \exp(- \tau / \lambda)$, where $\lambda=38$s is the average time interval.
Their distribution in total mass and redshift is represented in Fig. \ref{fig:summary_plot}. The total mass in the source frame ranges from \unit[$2.2$ to $202$]{M$_\odot$}, and redshift from $0.04$ to $13.96$.
Waveforms are generated using the \texttt{IMRPhenomPNRv2} approximant \cite{2021PhRvD.104l4027H} for BNS, and \texttt{IMRPhenomXPHM} \cite{2021PhRvD.103j4056P} for BBH and NSBH. Extrinsic parameters such as location on the sky, cosine of the inclination, polarization angle and phase at origin are randomly chosen following a uniform distribution. 
The dataset used for this MDC will be described in more details in a future document.

\subsection{Search methodology}

 We analyze this dataset with \texttt{PySTAMPAS}, a data analysis pipeline designed to search for un-modeled GW transients with duration ranging from few seconds to minutes. A detailed overview of the pipeline and its implementation can be found in \cite{PySTAMPAS}. The following describes its basic workflow and the configuration used for this analysis.

\paragraph*{Time-frequency maps}
The dataset is split into \unit[$512$]{s}-long analysis windows. For each window and each of the $3$ detectors E1, E2, and E3, a multi-resolution time-frequency map (\textit{tf-map}) of the data is built by computing the Fourier Transform over short segments of duration $0.5$, $1$, $2$, and \unit[$4$]{s}. These tf-maps span a frequency range between $4$ and \unit[$2000$]{Hz}, and each pixel contains the autopower SNR $\rho(t;f)$ that is the Fourier Transform divided by the square root of the Power Spectral Density (PSD) of the noise. \note{The PSD is estimated using a variant the Welch method that takes the median over Hann-windowed overlapping time segments \cite{FindChirp}, in order to mitigate the effect of high SNR signals present in the data.}

\paragraph*{Clustering}
GW signals typically appear in tf-maps as clusters of pixels that stand above the noise. In order to identify candidate GW events, a clustering algorithm is applied to each tf-map.
Different clustering strategies can be used to target different signal morphologies. For this analysis, we use two complementary algorithms.

The first one is a seed-based clustering algorithm, which identifies groups of bright pixels by time-frequency proximity to form clusters. Seed-based algorithms are sensitive to any signal morphology, provided that it is contiguous in the tf-map and loud enough to generate high SNR pixels. They are therefore suited for generic unmodeled searches with minimal assumptions on the morphology of the signals searched. We use the \texttt{burstegard} algorithm \cite{Pr2016}, that is commonly used for long-duration burst searches in LVK data \cite{2019PhRvD..99j4033A, 2021PhRvD.104j2001A}.

Complementary to seed-based clustering, \textit{seedless} algorithms  try to fit pre-determined templates (e.g Bezier curves) onto the tf-maps \cite{Seedless1, Seedless2}. The main advantage of seedless algorithms over seed-based ones is that they do not require bright pixels, and can therefore reconstruct fainter signals. However, they are only sensitive to signals whose morphology correspond to the class of templates tested. Long-duration signals with a locally-narrow frequency band, appearing as long tracks in tf-maps, are the primary targets of such algorithms; inspirals of stellar-mass CBC in a detector like ET belong to this class of signals.
Here we run a seedless clustering algorithm derived from \cite{2014PhRvD..90h3005C} that uses Newtonian chirps as template, in order to target specifically the BNS signals.

\paragraph*{Cross-correlation}
Clusters extracted from single-detector tf-maps are then cross-correlated with data from the two other detectors to build a coherent \textit{trigger} that is the final GW candidate. Each trigger is assigned a ranking statistic $p_{\Lambda}$ that reflect its significance. This statistic is based on the summed coherent SNR of the pixels and the residual energy in each detector (see \cite{PySTAMPAS} for more details on the construction of $p_{\Lambda}$). 
Since candidates are first extracted as clusters of pixels from single-detector tf-maps, a GW signal may be identified in several detectors and generate several triggers. Therefore, when triggers found in different detectors overlap by more than $90\%$ in time and frequency, only the most significant is kept.

\section{Results}
\label{sec:results}
\subsection{Background estimation}

To assess the significance of candidate events found by the search, one needs to estimate the rate of background triggers, i.e., triggers that are generated by noise fluctuations in the detectors. In general, GW detectors' noise is not Gaussian and presents many transient features that generate noise triggers whose distribution cannot be predicted analytically. Therefore, the background distribution must be estimated empirically.
As many other GW search algorithms, \texttt{PySTAMPAS} implements the time-slides method, which consists in shifting the data between the detectors by an amount of time greater than the maximal duration of the GW signal searched. Therefore, the properties of the noise in individual detectors are preserved, while any GW signal present in the data appears as incoherent.
This method also allows to simulate a large amount of background data by replicating the experiment with different values of the time shift. The cumulative rate of events found in time-shifted data is used as an estimator of the false-alarm rate (FAR) of the search.

\begin{figure}
    \centering
    \includegraphics[width=\columnwidth]{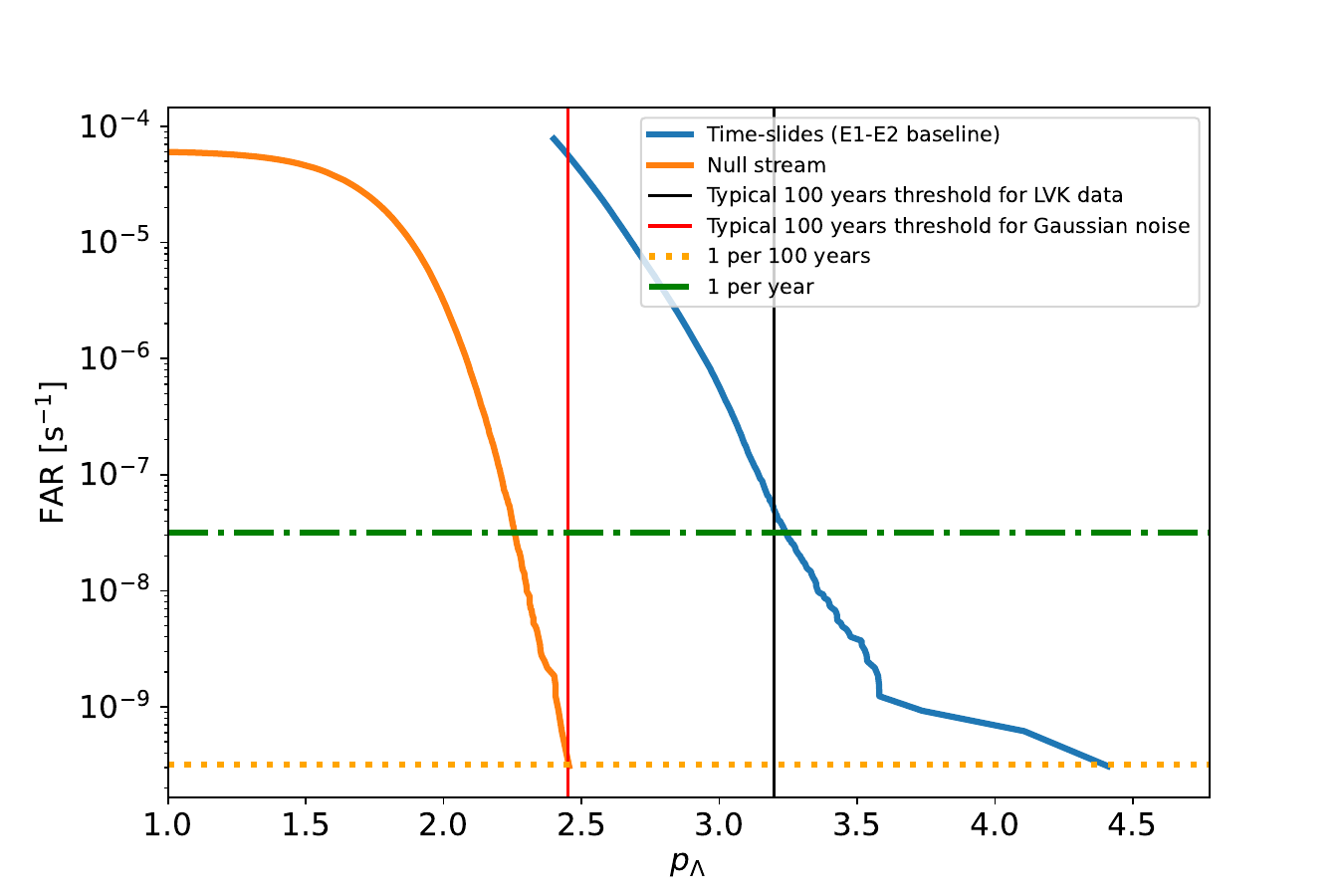}
    \caption{Cumulative rate of noise triggers as a function of the detection statistic $p_{\Lambda}$ for the two methods used to estimate the background.}
    \label{fig:FAR}
\end{figure}

\paragraph*{Time shifts}
In \texttt{PySTAMPAS}, clusters of pixels extracted from each detector are time-shifted with respect to the other detectors' data to estimate the background. For this search, we use $1216$ different values for the time shift, simulating $\sim 102$ years of background noise.
The cumulative rate of background triggers as a function of their ranking statistic $p_{\Lambda}$ is represented by the blue curve in Fig. \ref{fig:FAR} for the E1-E2 baseline. It is significantly higher than the rate we would expect for pure Gaussian noise, which is shown by the orange curve, and even higher than what we typically observe in realistic LVK searches, where the noise is dominated by loud transient artifacts \cite{PySTAMPAS}. 
Due to the high rate of high SNR signals in the dataset, most of the clusters extracted from single-detector tf-maps actually correspond to GW signals, and there is a significant probability that a signal from one detector is time-shifted on top of another signal in the other detectors, creating coherent pixels that increase the overall significance of the trigger. The main consequence is an apparent reduction of the sensitivity of the search, as the significance of events found in coincidence is reduced.





\paragraph*{Null stream}
This result illustrates the (well-known) limits of the time slides method when the data contains many loud signals, as it is expected for third-generation GW detectors such as ET. However, the triangular configuration considered here for ET offers a convenient way to bypass this issue in the form of the \textit{null stream}.
In general, the null stream is a linear combination of the output of the detectors that cancels a given GW signal.
Such combination always exists for an array of three or more detectors, as was first demonstrated by the authors of \cite{PhysRevD.40.3884}.
In the general case of arbitrarily located and oriented detectors, that combination depends on the sky position of the source, which must therefore be known \textit{a priori} (see e.g \cite{2006PhRvD..74h2005C} or \cite{2010NJPh...12e3034S} for the mathematical details on the computation of the null stream).
However, in the particular case of a triangular configuration, the sum of the responses of the three detectors to any GW signal is always zero \cite{2009CQGra..26h5012F, 2012PhRvD..86l2001R}. This allows one to build a single null stream that effectively cancels all GW signals present in the data, assuming the three interferometers are perfectly calibrated \cite{2020arXiv200910212S}.

We now use the null stream to estimate the distribution of noise triggers. As the noise used for this MDC is purely Gaussian, this amounts to estimating the background distribution on Gaussian noise.
The cumulative rate of background events obtained on the null stream is represented by the orange curve in Fig. \ref{fig:FAR}, and corresponds indeed to the distribution expected for pure Gaussian noise. We use it to set a threshold of on $p_\Lambda$ that corresponds to a FAR of $1$ per year.


\subsection{Search for generic transients with \texttt{burstegard}}

\begin{figure}
    \centering
    \includegraphics[width=\columnwidth]{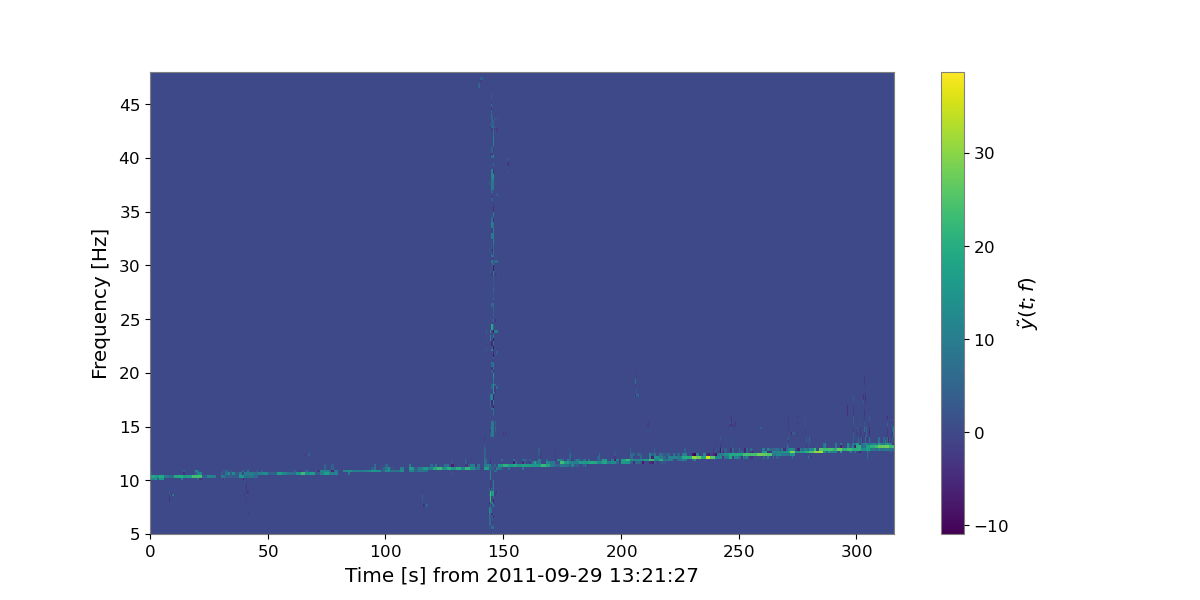}
    \includegraphics[width=\columnwidth]{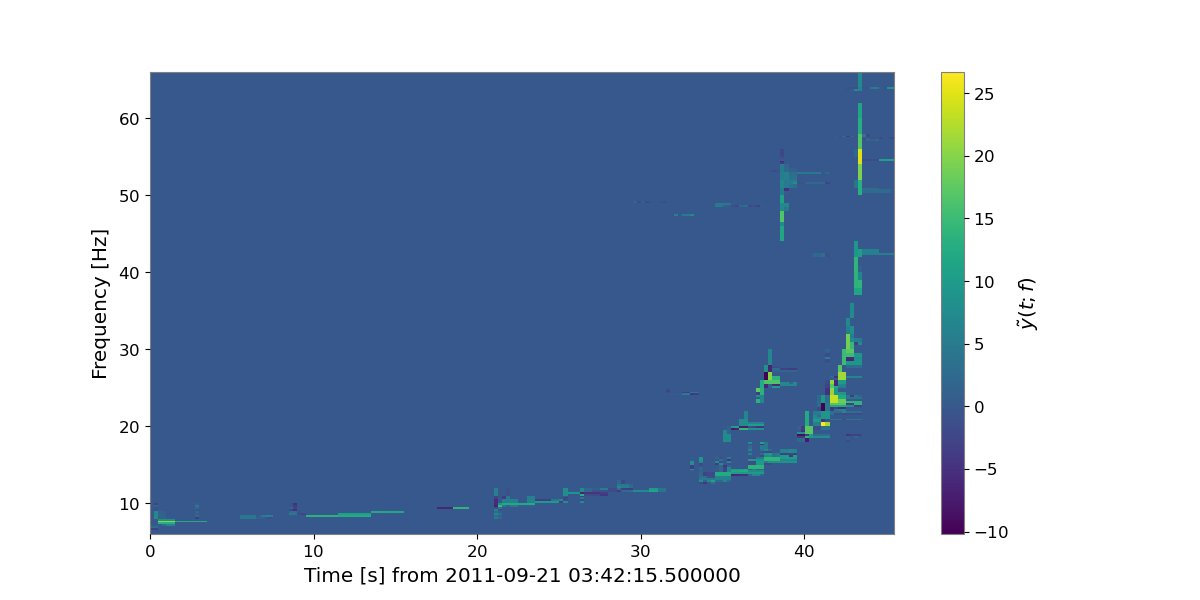}

    \caption{Time-frequency maps of significant triggers recovered by \texttt{PySTAMPAS} that are associated with an injection. In the top figure, the quasi-horizontal track between $10$ and \unit[$15$]{Hz} corresponds to a loud BNS signal, while the quasi-vertical one around \unit[$150$]{s} corresponds to a BBH signal. In the bottom figure, two different injected signals have been reconstructed as one trigger because they overlap partly.}
    \label{fig:ftmap_overlap}
\end{figure}

We analyze the coincident E1-E2-E3 data and identify $3065$ unique triggers with a FAR lower than $1$ per year. 
We use the following procedure to match triggers with injected signals, the list of which is provided by the authors of the MDC.
\begin{enumerate}
    \item For each trigger, we identify all injections for which the coalescence time lies within \unit[$20$]{s} of the end time of the trigger.
    \begin{itemize}
        \item If no injection exists inside this time window, the trigger is dismissed.
        \item If more than one injection meets this criterion, we associate the one which has the highest network SNR to the trigger.
    \end{itemize}
    \item For each injection associated with a trigger, we look at the number of triggers associated with it.
    \begin{itemize}
        \item If more than one trigger is associated with an injection, we keep the one that has the highest ranking statistic.
    \end{itemize}
\end{enumerate}
In the end, $2620$ out of $3035$ triggers ($85\%$) are uniquely matched with one injection.
Less than $1\%$ of the total number of significant triggers are not matched with any injection. This can happen for two reasons: first, because the analysis windows are only \unit[$512$]{s} long, a signal may be split between two or more windows and generate triggers in several windows. Because the matching procedure looks only at the coalescence time, triggers that correspond to the earlier parts of the signal will not be matched. 
\revision{In theory, it would be possible to increase the duration of analysis windows to be able to better reconstruct hours-long BNS signals, but that would be at the cost of an increased computing time, as the computational cost of the clustering algorithms used in this search increases with the number of pixels in the time-frequency map.}
Also, \texttt{burstegard} may cluster noise pixels at the end of a signal, leading to the end time of the trigger being overestimated, sometimes by more than the threshold of \unit[$20$]{s} chosen for matching.
We consider these reasons as intrinsic limitations of the \note{seed-based clustering algorithm}, and therefore we decide to discard these triggers even if they correspond to actual GW signals. These limitations are illustrated in Fig. \ref{fig:ftmap_overlap}, which shows examples of triggers extracted by \texttt{PySTAMPAS}: in both cases, two high-SNR signals have been reconstructed as a single event because they overlap in time and frequency. In the first case, because the merger time of the BNS is not located in the same analysis window, none of these two injections have been associated with the trigger.
Conversely, one injected signal can be matched with several triggers. Because of the seed-based nature of the clustering algorithm, a signal may be split into several clusters. If this split happens close to the merger time, the two triggers will both have end times close to the reference coalescence time and will then be associated with the same injection. 
\note{We note that in general in MDCs, the procedure to associate recovered triggers to injected signals must rely on at least some arbitrary choices. For this particular study, we have chosen to use the difference between the trigger end time and the coalescence time of the CBC as the main criterion, because the seed-based clustering algorithm tends to better recover the merger part of the CBC. We find that a tolerance window of \unit[$\pm20$]{s} is sufficiently short to suppress the risk of false association while minimizing the number of injections missed.}

\begin{figure*}
    \centering
    \includegraphics[width=2\columnwidth]{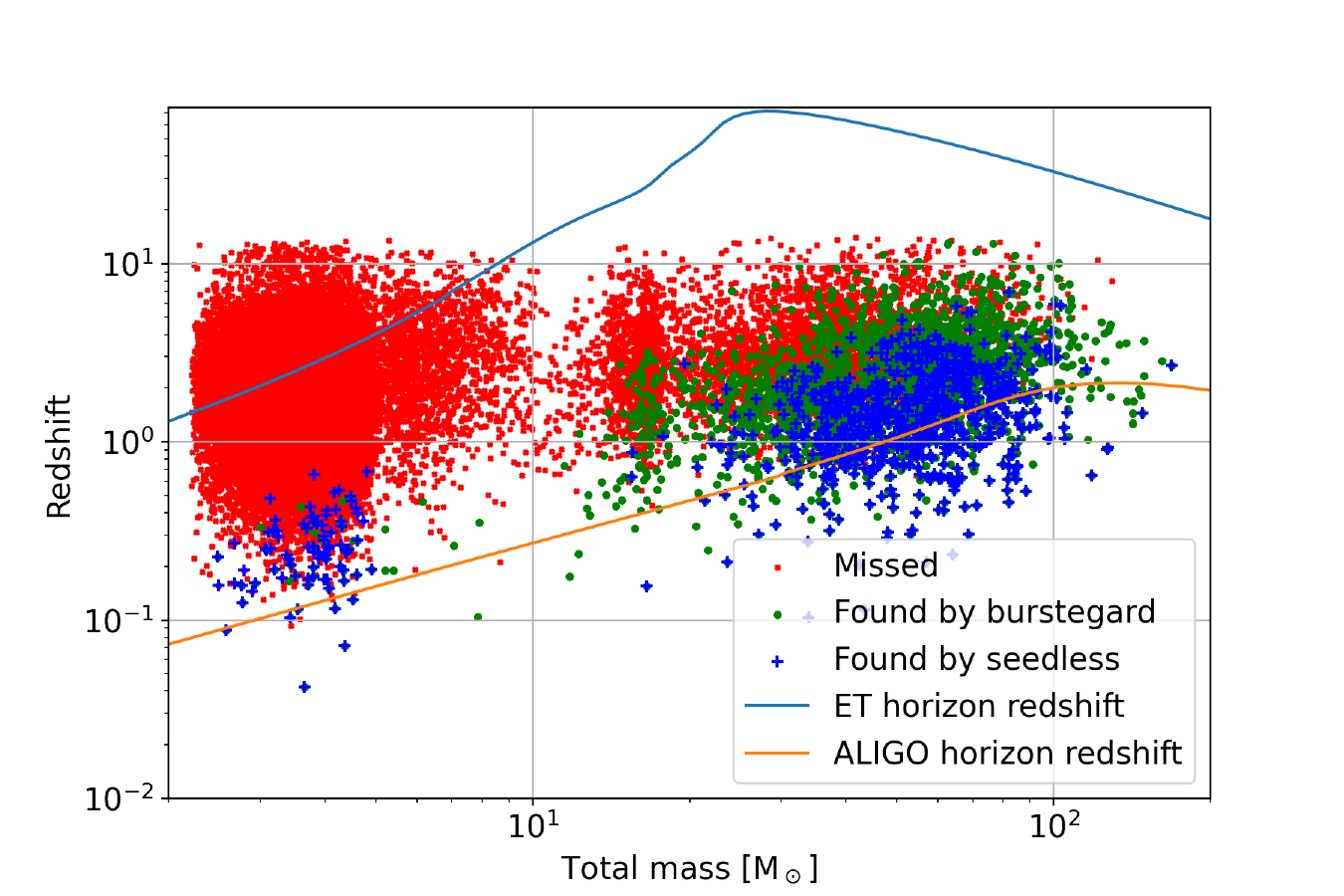}
    \caption{Redshift of injected CBC signals as a function of their total mass in the source frame. Green dots represent signals recovered using the seed-based  \texttt{burstegard} clustering algorithm, while blue crosses are signals recovered by the seedless algorithm. Red crosses represent missed injections. The blue and orange curves representr the maximum redshift at which an equal-mass, non-spinning, quasicircular binary coalescence would be detectable with an optimal SNR $\geq 8$ in ET and Advanced LIGO at design sensitivity respectively.}
    \label{fig:summary_plot}
\end{figure*}

Among the $2620$ injections recovered, $2567$ are labelled as BBH, $44$ as BNS, and $8$ as NSBH. The farthest event recovered is at a redshift $z=13.1$. It is a BBH with a total mass of \unit[$62.8$]{M$_\odot$} in the source frame (\unit[$885$]{M$_\odot$} in the detector frame). The distribution of recovered and missed signals in mass and redshift is shown in Fig. \ref{fig:summary_plot} for the two clustering algorithms used.
In total, this search recovers $38\%$ of the total number of injected BBH, including $65\%$ of the ones with a total mass in the source frame above \unit[$50$]{M$_\odot$}, and $89\%$ above \unit[$100$]{M$_\odot$}. High-mass BBH have a relatively short duration (of the order of a few seconds) in the frequency band of the detectors, so they are spread among a smaller number of time-frequency pixels than lower mass CBC, which are also less energetic. Therefore, they generate high SNR pixels that are picked up by the clustering algorithm, which explains why they are well recovered by \texttt{PySTAMPAS}, despite the pipeline being tuned to detect longer duration signals. 
It is therefore likely that other unmodeled search algorithms designed to target  signals with duration ranging from milliseconds to few seconds would be able to recover an even larger fraction of BBH.
\revisions{In LVK data, these high mass BBH have even shorter durations (less than $1$ s), and are therefore not detectable at all by the pipeline.}

\revisions{When using the time-slides method to estimate the background instead of the null stream, the threshold on the ranking statistic $p_\Lambda$ corresponding to a FAR of $1$ per year increases, as shown by the blue curve in Fig. \ref{fig:FAR}, which in turn reduces the number of signals effectively recovered from $2620$ to $1411$, a decrease of $46\%$ for the BBH signals.}


This search recovers $44$ BNS in $31$ days of data, corresponding to a rate of $\sim 1.4$ per day. The farthest BNS recovered is at a redshift $z=0.49$, while $50\%$ of the total injected BNS are recovered at $z<0.15$.
In addition to being intrinsically fainter than BBH, BNS signals are present in the frequency band of the detectors for a much longer time, so their total SNR is spread over a large number of pixels in the tf-map. Therefore, a seed-based clustering algorithm is not the best suited to recover such kind of signals, which typically generate long tracks of faint pixels. In the following paragraph, we test a different clustering algorithm aimed to improve the recovery of BNS signals.

\subsection{Search for chirp-like signals}
We use the seedless clustering algorithm described in \cite{2014PhRvD..90h3005C} to target long-duration, chirp-like tracks produced by BNS in tf-maps.
In this case, the time-frequency templates are only characterized by the chirp mass $\mathcal{M}_c$ and the coalescence time $t_c$:
\begin{equation}
    f(t) = \frac{1}{\pi} \left(\frac{5}{256}\right)^{3/8} \left( \frac{G \mathcal{M}_c}{c^3} \right)^{-5/8} (t_c - t)^{-3/8}.
\end{equation}
In reality, the actual GW signal of a BNS depends on more parameters than the chirp mass (such as individual component masses, spins and tidal deformabilities), and is more accurately described with post-Newtonian terms. Yet, our approximation allows us to limit the size of the template bank, hence the computational cost of the search, and should be sufficient for a basic search based on excess-power in tf-maps.

For each tf-map, we test $100$ values of $\mathcal{M}_c$ logarithmically spaced between $1$ and $5$ M$_\odot$, and $1024$ values of $t_c$ separated by \unit[$0.5$]{s}. \revision{We also run the same search with $100$ values of $\mathcal{M}_c$ spaced between $5$ and $200$ M$_\odot$ to determine whether this algorithm is also sensitive to BBH and NSBH.}
For each template, the corresponding cluster is the set of pixels $\Gamma$ that overlap the graph $(t, f(t))$. We define the (single-detector) ranking statistic
\begin{equation}
    \Lambda = \frac{1}{\sqrt{N}} \sum\limits_{(t;f) \in \Gamma} \rho(t;f)
\end{equation}
where $N$ is the number of pixels in the cluster.
The cluster with the highest value of $\Lambda$ in a tf-map is kept, and the corresponding values of $\mathcal{M}_c$ and $t_c$ are saved.
As shown in \cite{2014PhRvD..90h3005C}, the process is easily parallelizable so testing $\sim 10^5$ templates is computationally inexpensive (processing a \unit[$512$]{s}-long ft-map takes $\mathcal{O}(\unit[10]{s})$ on an off-the-shelf modern CPU).

\begin{figure}
    \centering
    \includegraphics[width=\columnwidth]{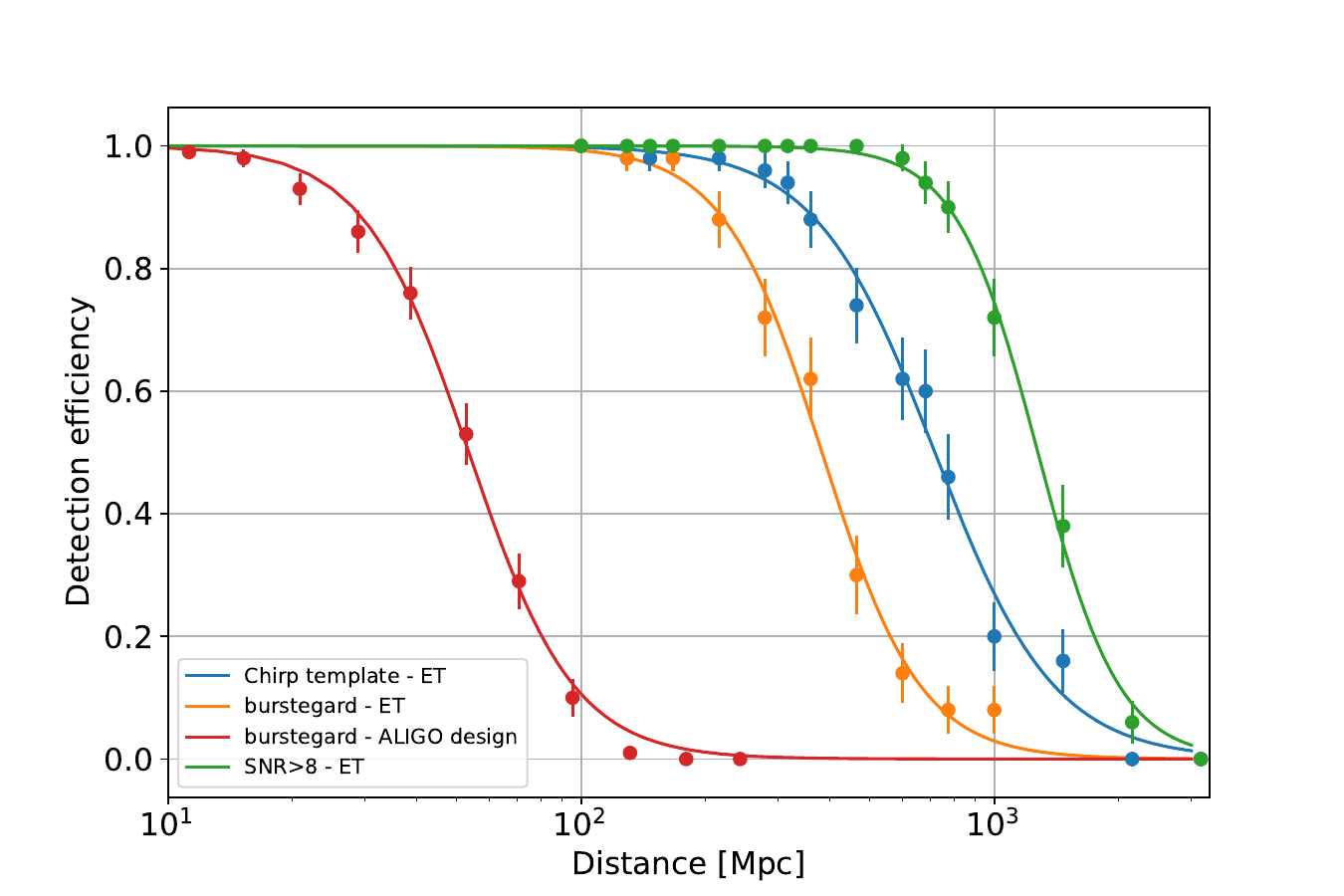}
    \caption{Detection efficiency as a function of the distance for a \unit[$1.4-1.4$]{M$_\odot$} BNS (IMRPhenomPv2 approximant) with seed-based (orange) and seedless (blue) clustering. The green curve represents the fraction of signals for which the matched filter SNR is above $8$ in at least one detector. The red curve represents the detection efficiency for the same waveform on Gaussian noise following Advanced LIGO design sensitivity for comparison (only the .}
    \label{fig:BNS_range}
\end{figure}

In order to test the algorithm in a controlled way, we estimate the detection efficiency of the search for a reference \unit[$1.4-1.4$]{M$_\odot$} non-spinning BNS on the noise used for this MDC and compare the two clustering algorithms. 
The detection efficiency as a function of the injected distance is shown in Fig. \ref{fig:BNS_range} for a FAR of $1$ per year. The horizon, defined as the distance for which the detection efficiency reaches $50\%$, is \unit[$382$]{Mpc} for \texttt{burstegard}, and goes up to \unit[$717$]{Mpc} when using this implementation of seedless clustering, an increase of $88\%$. The distance at which $50\%$ of events have an SNR $\geq 8$ in at least one detector is \unit[$1275$]{Mpc}.
\revisions{As a comparison, we do the same study on Gaussian noise following Advanced LIGO design sensitivity and find a horizon distance of \unit[$53$]{Mpc} using \texttt{burstegard}, a decrease by a factor $\sim 7$ which is consistent with the relative sensitivities of the detectors.}

On the MDC data, we recover $81$ BNS in total, $84\%$ more than with \texttt{burstegard}. The farthest BNS recovered is at redshift $z=0.68$, and $50\%$ of BNS below $z=0.17$ are recovered. \revision{However, we recover only $712$ BBH and NSBH, $72\%$ less than \texttt{burstegard}, which is expected because seed-based clustering performs better on shorter signals.} The distribution of CBC recovered using the seedless clustering algorithm is shown by the blue crosses in Fig. \ref{fig:summary_plot}.
\revisions{When the background is estimated with the time-slides method, only $15$ BNS are recovered, a decrease of $81\%$. As explained above, BNS typically generate fainter pixels than BBH in time-frequency maps, so the associated triggers have a lower ranking statistic, leading to more of them being discarded when the recovery threshold increases.}

\begin{figure}
    \centering
    \includegraphics[width=\columnwidth]{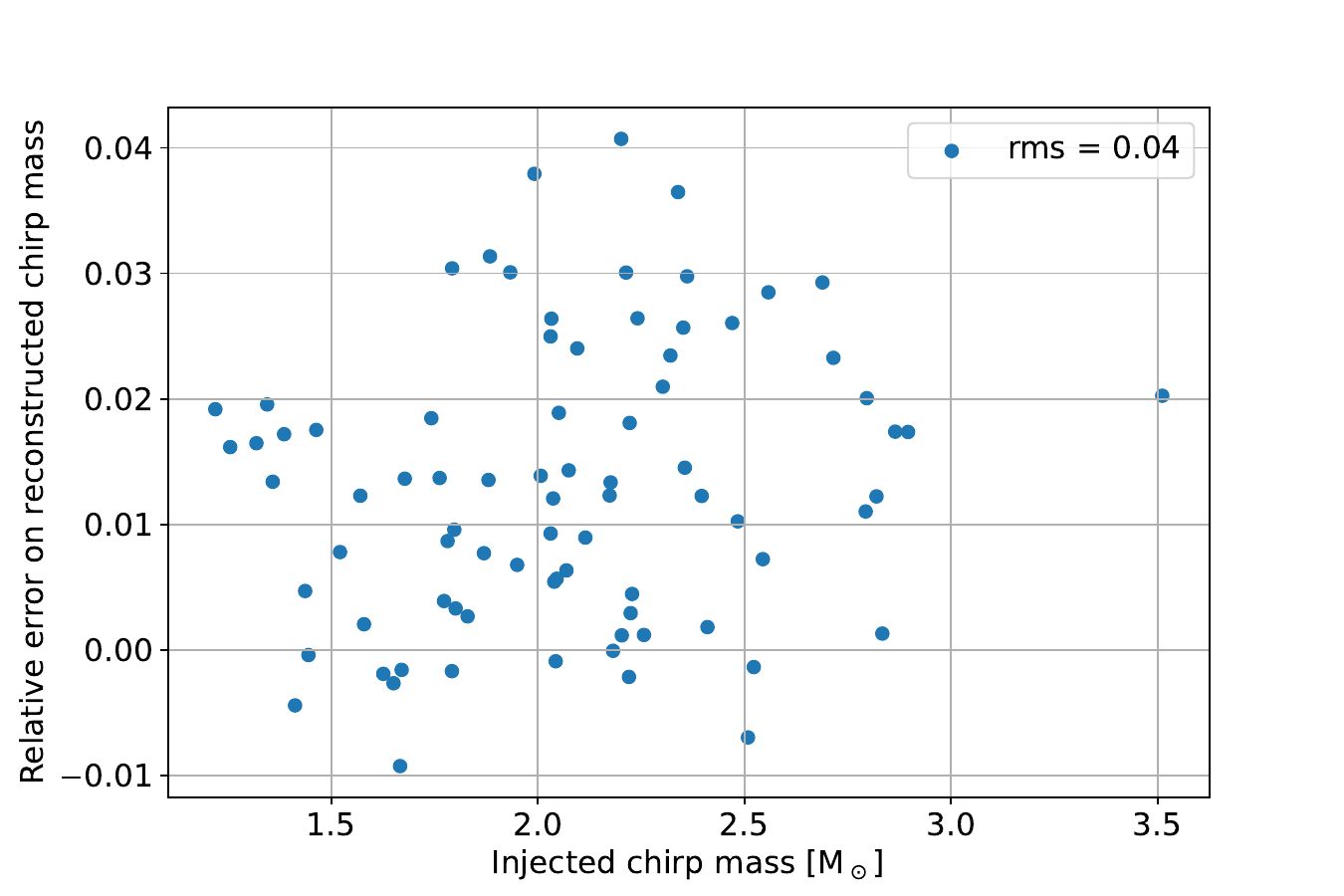}
    \includegraphics[width=\columnwidth]{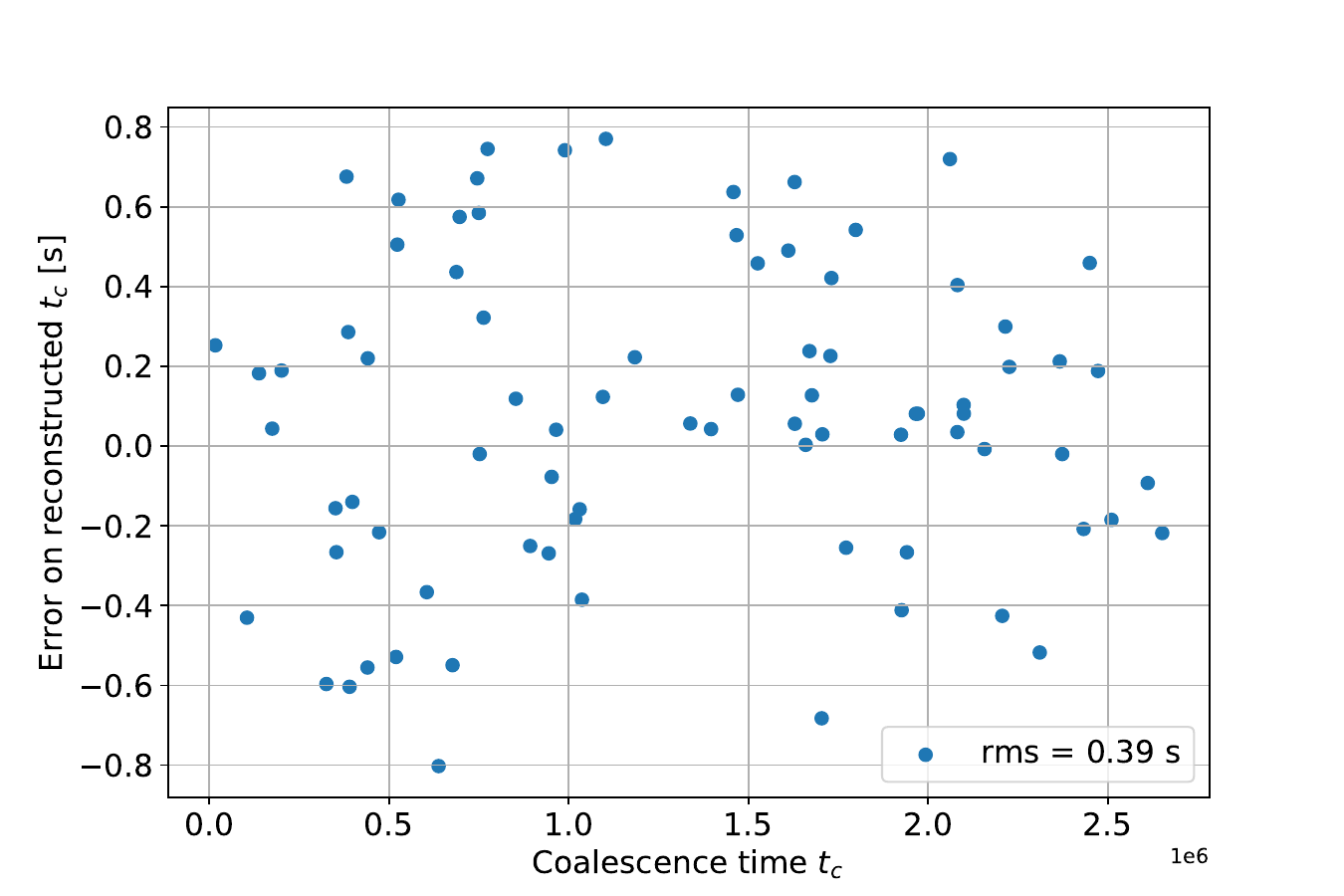}
    \caption{Relative error on the reconstructed chirp mass in the detector frame (top) and on the reconstructed coalescence time (bottom) as a function of the injected one for the $81$ BNS signals recovered by the search using seedless clustering.}
    \label{fig:MC_rms}
\end{figure}

Finally, we compare the chirp mass \revision{and coalescence time} recovered by the algorithm to the values injected for each recovered signal in Fig.~\ref{fig:MC_rms}.
\revision{Given that the algorithm performs poorly for BBH and NSBH, we focus here on the search targeting chirp masses between $1$ and $5$ M$_\odot$.}
The root mean square error (RMS) between the two values is \unit[$0.04$]{M$_\odot$} for BNS, which corresponds to an average relative error of $1.3\%$. 
We note that the chirp mass estimated here is the one in the detector frame. To estimate the mass in the source frame, an independent estimation of the redshift would be needed, which is outside of the scope of this method and would probably have to rely on full Bayesian inference with CBC waveform models triggered by the \texttt{PySTAMPAS} candidate \cite{Bilby}.
The RMS for the coalescence time is \unit[$0.39$]{s}, which is consistent with the \unit[$0.5$]{s} separation between the different values of $t_c$ tested, \revision{but does not match the precision of $\sim 0.1$ s reached by matched-filtering pipelines.}


\section{Effect of the CBC foreground on the detectability of other transient signals}
\label{sec:unmodeled}
An important issue regarding ET data analysis is the effect of the CBC foreground on the detection capabilities of other GW sources that are expected to be present in the data.
In the case of stochastic GW backgrounds, it has been shown that the CBC foreground will indeed prevent the observation of a stochastic background of cosmological origin, which motivates the development of methods to accurately substract CBC signals from the data \cite{2008PhRvD..77l3010H, 2021PhRvD.103d3023M, 2020PhRvL.125x1101B, 2020PhRvD.102b4051S}.
Here, we study how CBC events recovered by \texttt{PySTAMPAS} affect the ability of the pipeline to detect other kind of unmodeled signals. In order to do that, we inject a selection of waveforms routinely used to characterize unmodeled long-duration GW searches on top of the MDC data, and on top of pure Gaussian noise following the same sensitivity curve as the MDC noise, and compare their recovery using \texttt{burstegard}.

We select $4$ waveforms that span the parameter space of long-duration transients in frequency range, duration, and spectral morphology. They come from models of accretion disk instabilities around a black hole (ADI-A) \cite{ADI}, eccentric CBC (ECBC-C) \cite{2018PhRvD..97b4031H}, and post-merger magnetars (magnetar-D) \cite{2015ApJ...798...25D}. We also use a quasi-monochromatic Sine Gaussian waveform with a central frequency of \unit[$405$]{Hz} and a decay time of \unit[$50$]{s} (SG-C).

For each waveform, we select $10$ logarithmically spaced distances. For each distance, we inject $100$ signals with random sky position, polarization angle, and cosine of the inclination. A signal is considered to be recovered if there is a trigger within the time and frequency limits of the injection with a FAR lower than $1$ per year, and if there is no trigger that meets these conditions in the data without the injection. Therefore,  in the case where an unmodeled injection and a loud CBC event overlap, the unmodeled injection is rejected.
We use the fraction of recovered signals at a given distance as an estimator for the detection efficiency. The final figure of merit for a given waveform is the distance at which detection efficiency reaches $50$ \%. These values are reported in Table \ref{tab:longdur_wvf} for both cases.

\begin{table}[]
    \centering
    \begin{tabular}{|c|c|c|cc|}
    \hline
        \multirow{2}{*}{Waveform} & \multirow{2}{*}{Frequency [Hz]} & \multirow{2}{*}{Duration [s]} &  \multicolumn{2}{c|}{Distance [Mpc]} \\
        &   &   &  Noise & MDC data \\
         \hline
         ADI-A & $135-166$ & $39$ & $445$ & $425$\\
         ECBC-C & $10-300$ & $297$ & $312$ & $301$ \\
         magnetar-D & $1598-1900$ & $400$ & $2.80$ & $2.80$ \\
         SG-C & $402-408$ & $250$ & $4.57$ & $4.49$ \\

         \hline
    \end{tabular}
    \caption{Distance at $50\%$ detection efficiency for a FAR of $1$ per year for a set of standard long-duration GW waveforms.}
    \label{tab:longdur_wvf}
\end{table}

In average, the horizon distance is reduced by $3.5 \%$ in the case where CBC signals are present in the data. Indeed, a fraction of about $1\%$ of triggers are rejected because they overlap with CBC triggers, but it is small enough to not impact significantly the sensitivity of the search. This results is dependant on the frequency of the signal: waveforms with central frequency below a few hundred \unit{Hz}, like ADI-A and ECBC-C, are slightly more affected by the presence of CBC than waveforms at higher frequencies.
Hence, we show that the presence of many detectable CBC only marginally affects the detection efficiency of long-duration burst signals. 
This result is in line with and complements previous studies that showed that the confusion noise created by overlapping CBC signals does not significantly affect the detection of CBC using matched filtering techniques \cite{2012PhRvD..86l2001R, 2023PhRvD.107f3022W, 2024PhRvD.109h4015J}.

\section{Conclusion}
\label{sec:conclusion}

We have analyzed a mock data challenge consisting of one month of data simulating a triangular Eintein Telescope and a realistic distribution of CBC signals with \texttt{PySTAMPAS}, a data analysis pipeline to search for unmodeled, long-duration GW signals. 
The main conclusions of this work are summarized in the following points.

\begin{itemize}
    \item Regarding background estimation, the time-slides method tends to greatly overestimate the rate of background triggers, because the fraction of overlapping signals in time-shifted data is no longer negligible compared to second-generation detectors. The null stream provides a convenient way to overcome this issue, but it is only available for specific geometrical configurations such as a triangle, and in practice would be affected by calibration errors and non-Gaussian noise. Future, more realistic MDCs will be necessary to properly assess if the null stream can really be used as a replacement for the time-slide method. \revisions{In this search, using the null stream to estimate the background instead of the time-slides method allows to recover $86\%$ more signals at a fixed FAR threshold}.
    
    \item This MDC highlights a few limitations of \texttt{PySTAMPAS} in the high signal regime, such as overlapping signals that are often reconstructed as a single trigger, or conversely signals that are split into several triggers. Clustering algorithms should be improved to overcome these issues, for example by using machine learning techniques to identify clusters made of several signals. One could also compare the relative SNR in different detectors to discriminate between overlapping signals originating from different parts of the sky, using the fact that they would be weighted differently by the antenna patterns of the detectors.

    \item \texttt{PySTAMPAS} is able to detect a large fraction of the total BBH in the universe, especially at high masses. Although it is not a dedicated CBC pipeline and therefore remains less sensitive than matched filtering techniques, the comparatively much lower computational cost could make it interesting for low-latency searches and independent confirmation of detections. It should also be more robust to ``exotic'' CBC signals whose parameters do not fall within the template bank of matched-filtering-based pipelines, such as systems with high eccentricities, matter effects or strong higher-order modes, as well as to uncertainties in the waveforms.
    \revisions{The better sensitivity of ET at low frequency increases the effective duration of BBH signals, which drives them in the parameter space covered by this search algorithm.}

    \item Using a dedicated seedless clustering algorithm that tries to fit Newtonian chirps onto the tf-maps, the search is able to recover a larger amount of BNS, up to $z=0.68$, with a detection rate of around $3$ per day, and to estimate the chirp mass and coalescence time with reasonable accuracy ($\sim 1 \%$) given the simplicity of the approach. This might be sufficient to trigger full Bayesian inference on the candidates, for more precise and accurate results. \revision{This approach was not designed for short duration signals and performs poorly on NSBH and BBH as expected.}

    \item The presence of the CBC foreground does not significantly affects the ability of the pipeline to detect long-duration burst signals, which are its primary target. 
    
\end{itemize}

We have therefore shown that \revision{\texttt{PySTAMPAS}} would be able to detect a significant fraction of CBC signals present in ET, and to provide \revision{a much quicker estimation of the chirp mass of BNS than modeled pipelines, with an average precision of 1.3\% as a byproduct of the search algorithm.}

The whole search, including the analysis of $102$ years of background, took $\sim 20$ hours using $120$ virtualized off-the-shelf x86\_64 CPU cores, which is much lower than what would likely be required by template-based searches \cite{Lenon2021}.
We stress that this is an analysis of one-month of ET data that can be carried out \emph{today} with commonplace computing resources.
The relatively low computational cost of this search could make it interesting for a quick look of ET data, for early-warning BNS alerts, and for an independent validation of detections in a context where a large number of high-SNR signals are present in the data and may not be easy to model with a sufficient accuracy for deep matched-filter searches.

\acknowledgments

Part of our simulations were performed on the Virtual Data cloud computing system at IJCLab.
We thank Michel Jouvin and Gerard Marchal-Duval for their prompt support and advice about this system.

\bibliographystyle{apsrev4-2}
\bibliography{references}

\end{document}